\documentclass[aps,pra,amsmath,amssymb,twocolumn,floatfix,nofootinbib,superscriptaddress,showpacs]{revtex4}
\usepackage[dvips]{graphicx}
\newcommand{\bra}[1]{\langle #1|}
\newcommand{\ket}[1]{|#1\rangle}
\begin{document}

\title{Bose-Fermi Pairs in a Mixture and the Luttinger Theorem within
  a Nozi\`eres-Schmitt-Rink like Approach}

\author{T. Sogo}
\affiliation{Institut de Physique Nucl\'eaire, CNRS-IN2P3 and
  Universit\'e Paris-Sud, F-91406 Orsay Cedex, France}

\author{P. Schuck}
\affiliation{Institut de Physique Nucl\'eaire, CNRS-IN2P3 and
  Universit\'e Paris-Sud, F-91406 Orsay Cedex, France}
\affiliation{Laboratoire de Physique et Mod\'elisation des
  Milieux Condens\'es, CNRS and Universit\'e Joseph Fourier, 25 Avenue
  des Martyrs, Bo\^ite Postale 166, F-38042 Grenoble Cedex 9, France}

\author{M. Urban}
\affiliation{Institut de Physique Nucl\'eaire, CNRS-IN2P3 and
  Universit\'e Paris-Sud, F-91406 Orsay Cedex, France}

\begin{abstract}
Boson-fermion pair correlations in a mixture are considered at zero
temperature in the $T$-matrix approximation. Special attention is
paid to the Luttinger theorem. In a strict RPA variant of the
Nozi\`eres-Schmitt-Rink approach, it is shown that this theorem is
respected also in the homogeneous infinite matter case. We calculate
the corresponding occupation numbers of fermions and bosons and the
condensate depletion. We also show that in the limit of very small
boson density, our results are in good agreement with the results
found in the literature for the Fermi polaron in strongly imbalanced
Fermi-Fermi mixtures.
\end{abstract}
\maketitle

\section{Introduction}
Cold atom physics is constantly progressing at a rapid pace. Fermi and
Bose systems have been under consideration extensively. Bose-Fermi
(BF) mixtures have been studied so far a little less. In early
attempts to create a degenerate Fermi gas, bosonic $^7$Li
\cite{Truscott2001,Schreck2001} or $^{23}$Na atoms
\cite{Hadzibabic2002} were added to the fermionic $^6$Li in order to
allow for sympathetic cooling. The first BF mixture with an attractive
BF interaction was that of $^{40}$K and $^{87}$Rb \cite{Roati2002}. In
present-day experiments with $^6$Li, a small fraction of $^7$Li atoms is
kept to serve as a thermometer \cite{Nascimbene2010}. In \cite{Wu2011}
a mixture of $^{40}$K, $^{41}$K, and $^6$Li was created with the main
goal to produce a mass-imbalanced Fermi gas of $^{40}$K and $^6$Li,
the boson $^{41}$K acting again as a coolant. The possibility to
produce a dipolar Fermi gas of polar fermionic molecules has triggered
many experiments with different BF mixtures such as $^{87}$Rb-$^{40}$K
\cite{NiOspelkaus2008,Cumby2013}, $^{23}$Na-$^{40}$K
\cite{ParkWu2012,WuPark2012}, and $^{23}$Na-$^6$Li
\cite{HeoWang2012}. Experiments with BF mixtures are not restricted to
alkaline atoms. For instance, also $^{84}$Sr-$^{87}$Sr \cite{Tey2010}
mixtures were created. 

From the theory perspective, BF mixtures are interesting in their own
right, e.g., to study the interplay between different quantum
statistics in various fields of physics. They also may serve to
simulate definite physical systems. For example, BF mixtures have been
considered as an analogy to what might happen in the quark-hadron
phase transition \cite{Maeda2009} within the scenario that first two
quarks form a tightly bound diquark (boson) which then combines with a
third quark (fermion) to form a nucleon. It, thus, is important to
further develop the theory of BF correlations in BF mixtures.  A
particularly interesting question concerns the structure and behavior
of BF pairs. In \cite{Storozhenko2005} we have shown that similar to
the formation of Cooper pairs in two component Fermi systems, also in
BF mixtures stable BF pairs can exist with very weak attraction for
which a bound state cannot be formed in free space.

In this work, we shall be concerned with bosons and fermions
interacting via a broad Feshbach resonance. Under this condition, the
system can be described by a Hamiltonian of bosons and (spinless)
fermions interacting via an attractive (or repulsive) contact
potential. There exist several Monte Carlo investigations in 1D
\cite{Pollet2006} and 3D \cite{Yamamoto2012,Bertaina2013} BF
systems. However, also approximate many body approches have been
applied. Among those several works using the so-called $T$-matrix
approximation have appeared
\cite{Storozhenko2005,Watanabe2008,Fratini2010,Ludwig2011,Fratini2012,Fratini2013}
and this shall also be our framework in this paper. The BF $T$-matrix
describes BF scattering states but also eventuel formation of bound
states. Bound states in the medium are especially interesting. The
$T$-matrix also can serve to study single particle properties. In this
respect, folding the $T$-matrix with either a fermion or a boson
propagator yields the boson or fermion self-energy of the Dyson
equation.

The $T$-matrix approximation has become particularly popular since
Nozi\`eres and Schmitt-Rink (NSR) showed that for attractive Fermi
systems this approach interpolates beween the weak coupling (BCS)
situation and the Bose-Einstein condensation (BEC) of strongly bound
fermion pairs \cite{NSR}. This approach has also been generalized to
study the pairing properties of polarized Fermi systems where there
exist more fermions with, e.g., spin $\uparrow$ than those with spin
$\downarrow$
\cite{LiuHu2006,Parish2007,Kashimura2012a,Kashimura2012b}. However,
these studies have revealed that in this case the standard NSR
approach may lead, in some regions of the parameter space, to
pathological results. A special case of particular interest is that of
an extremely imbalanced mixture, which can be treated by considering a
single atom of the minority species, the so-called polaron limit. The
case of Fermi polarons, i.e., a single fermion of spin $\downarrow$ in
a system of fermions with spin $\uparrow$, has been intensively
studied using a variational ansatz \cite{Chevy2006,Punk2009} and,
equivalently, a $T$-matrix approach \cite{Combescot2007,Baarsma2012},
as well as using a diagrammatic Monte-Carlo technique
\cite{Prokofev2008a,Prokofev2008b,Vlietinck2013}. All these results
can be directly applied to BF mixtures with a very small number of
bosons, because if one considers only a single impurity it does not
matter whether it is a fermion or a boson.

In the present paper, we will pay special attention to an aspect that
so far has not been considered, namely the Luttinger theorem
\cite{Luttinger1960}. This theorem states that the volume of the Fermi
sphere is not changed by interactions, or in other words, that the
reduction of the occupation numbers $\rho_{k<k_F}$ inside the Fermi
sphere is exactly compensated by the non-vanishing occupation numbers
$\rho_{k>k_F}$ outside the Fermi sphere. It is highly non-trivial to
respect this theorem within a non-perturbative approximation
scheme. Here, we will use a variant of the NSR approach adapted to BF
systems. A particularity of the NSR approach is that it treats the
self-energy in the single-particle Dyson equation only to first
order. This, for instance, means that the NSR approach, if suitably
adapted, is strictly equivalent to the Random Phase Approximation
(RPA), here in the so-called particle-particle (pp) channel which sums
pp and hh (hole-hole) ladders simultaneously \cite{RS}. The fact that
pp-RPA satisfies, among other things, the analog of the Luttinger
theorem in a system with a discrete level structure such as atomic
nuclei has been known for many years \cite{RS,Bouyssy1974}. It has
also been demonstrated for a BF system on a lattice
\cite{Barillier2008}. But to the best of our knowledge, this has never
been explicitly shown in a continuum case. It will be one of the
results of the present work to show this for an attractively
interacting infinite BF system.

The paper is organized as follows. In Section \ref{sec:2p} we discuss
the BF scattering in a BF mixture within the pp-RPA framework. In
Section \ref{sec:1p} we discuss the correlation effects on the
ground-state properties. Finally, in Section \ref{sec:summary} we
summarize and conclude.

\section{The boson-fermion $T$-matrix within particle-particle RPA}
\label{sec:2p}
The starting point of our study is the following BF hamiltonian:
\begin{multline}
\label{H}
H=
\int d^3r 
\Big[
-\psi^\dagger({\bf r})
\frac{{\bf \nabla}^2}{2m_F}
\psi({\bf r})
-
\varphi^\dagger({\bf r})
\frac{{\bf \nabla}^2}{2m_B}
\varphi({\bf r}) \\
+
g
\psi^\dagger({\bf r})
(\sqrt{n_0}+\varphi^\dagger({\bf r}))
(\sqrt{n_0}+\varphi({\bf r}))
\psi({\bf r})
\Big]
\end{multline}
where $\psi$ and $\varphi$ are the fermion and boson field operators,
$m_{F,B}$ are the fermion and boson masses, and $g$ is the coupling
constant. The field operator $\varphi$ has been shifted by a c-number
$\sqrt{n_0}$, where $n_0$ denotes the density of condensed bosons
\cite{FW}. The field operators $\psi$ and $\varphi$ can be written in
terms of fermion and boson annihilation operators $c_{\bf k}$ and
$b_{\bf k}$ as
\begin{gather}
\psi({\bf r}) 
= 
\int\frac{d^3k}{(2\pi)^3} c_{{\bf k}}e^{i{\bf k}\cdot{\bf r}}\,,
\\
\varphi({\bf r})
=
\int\frac{d^3k}{(2\pi)^3} b_{{\bf k}}e^{i{\bf k}\cdot{\bf r}}\,.
\end{gather}
Analogously, $\psi^\dagger$ and $\varphi^\dagger$ can be written in
terms of fermion and boson creation operators $c^\dagger_{\bf k}$ and
$b^\dagger_{\bf{k}}$.

The hamiltonian (\ref{H}) is suitable for the case of a broad Feshbach
resonance in the BF interaction \cite{Fratini2012}. We neglect the
boson-boson (BB) interaction. A repulsive BB interaction would
essentially result in a mean-field shift that can be absorbed in a
redefinition of the boson chemical potential and does not change the
results very much \cite{Watanabe2008,Fratini2010}. Since we assume
that the Fermions are present in only one spin state (``spinless
fermions''), there cannot be an $s$-wave fermion-fermion (FF)
interaction and higher partial waves are usually negligible in
ultracold trapped atoms.

As mentioned before, we want to apply a suitably adapted NSR approach
to the present BF problem. We will work at zero temperature and with
chronological Green's functions (GFs). In standard notation \cite{FW}
we have for the single-particle fermion and non-condensed boson GFs
\begin{gather}
G_F({\bf k},t-t')= -i\bra{0}T c_{{\bf k}}(t)c^{\dagger}_{{\bf k}}(t')\ket{0}\,,\\
G_B({\bf k},t-t')= -i\bra{0}T b_{{\bf k}}(t)b^{\dagger}_{{\bf k}}(t')\ket{0}\,,
\end{gather}
where $T$ means time-ordering. The corresponding free boson and
fermion propagators in frequency space are given by
\begin{equation}
G_{B}^{0}({\bf k},\omega)
=
\frac{1}{\omega-\varepsilon_B(k)+i\eta},
\end{equation}
and
\begin{equation}
G_{F}^{0}({\bf k},\omega)
=
\frac{\theta(k-k_F)}{\omega-\varepsilon_F(k)+i\eta}
+
\frac{\theta(k_F-k)}{\omega-\varepsilon_F(k)-i\eta}\,,
\label{Gfermifree}
\end{equation}
where $\varepsilon_{B,F}(k)=k^2/(2m_{B,F})$ are the non-interacting
boson and fermion energies and $k_F$ is the Fermi momentum, related
to the fermion density $n_F$ by $n_F = k_F^3/(6\pi^2)$.

\begin{figure}
\includegraphics[width=80mm]{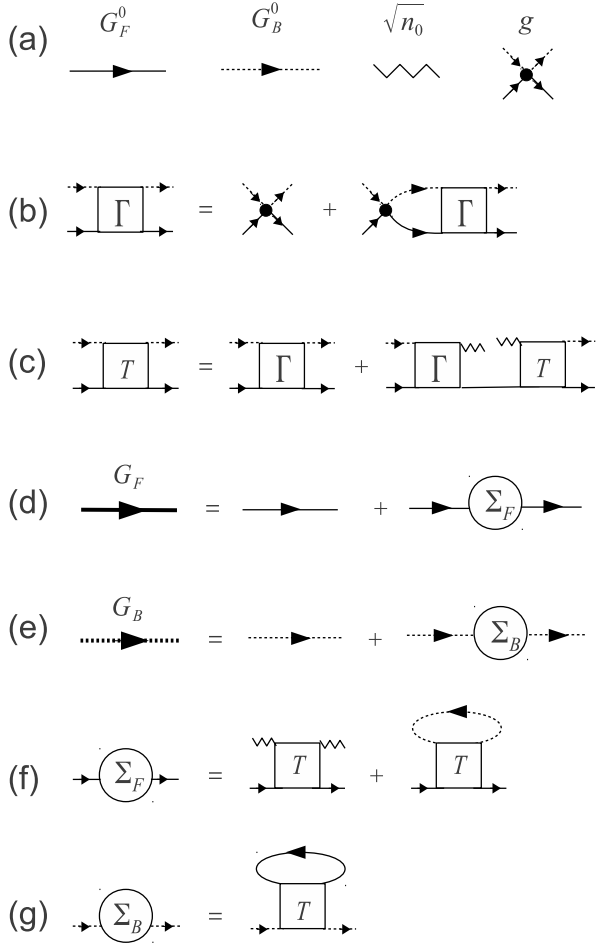}
\caption{\label{fig-feynmandiagram} Feynman diagrams corresponding to
  the formulas in the text.}
\end{figure}

We use these free GFs to construct the BF $T$-matrix in ladder
approximation. The result can be written as
\cite{Storozhenko2005,Watanabe2008}
\begin{equation}
T({\bf k},\omega)
=
\frac{1}{\Gamma^{-1}({\bf k},\omega)-n_0G_F^0({\bf k},\omega)}.
\label{tmatrix}
\end{equation}
The regularized BF scattering matrix $\Gamma$ with no boson in the
condensate is a standard expression which can be found in the
literature \cite{Watanabe2008,Fratini2010,Fratini2012}
\begin{eqnarray}
\Gamma({\bf k},\omega)
=
\frac{1}{\frac{m_r}{2\pi a}-J({\bf k},\omega)}
\end{eqnarray}
where $m_r=m_Fm_B/(m_F+m_B)$ is the reduced mass, $a$ is the BF
scattering length, and $J$ denotes the uncorrelated BF propagator that
is given by
\begin{equation}
J({\bf k},\omega)
=
\int \frac{d^3k'}{(2\pi)^3}
\left[
\frac
{1-\theta(k_F-\left|\frac{m_F}{M}{\bf k}+{\bf k}'\right|)}
{\omega-\frac{k^2}{2M}-\frac{k^{\prime 2}}{2m_r}+i\eta}
+\frac{2m_r}{{k'}^2}
\right]\,,
\end{equation}
with $M=m_B+m_F$. The Feynman diagrams corresponding to the $\Gamma$
and $T$-matrices are shown in Fig.~\ref{fig-feynmandiagram}(b) and
(c).

For the following calculations it is important to study the analytical
properties of the $T$-matrix. The continuum threshold, i.e., the
energy above which the $T$-matrix has an imaginary part, lies at
\begin{equation}
\omega_{th}(k)
=
\left\{
\begin{array}{cc}
\frac{(k-k_F)^2}{2m_B}+E_F
&
\left(k \leq \frac{M}{m_F}k_F\right)
\vspace{5mm}\\
\frac{k^2}{2M}+E_F
&
\left(k > \frac{M}{m_F}k_F\right)
\end{array}
\right.
\end{equation}
where $E_F=k_F^2/(2m_F)$. For not too high momenta $k$,
$\Gamma(k,\omega)$ has a pole at $\Omega_\Gamma(k)$ below this
threshold. As a consequence, the $T$-matrix has one or two poles below
threshold:
\begin{align}
T({\bf k},\omega) 
=& 
\frac{\omega-\varepsilon_F(k)}
{(\omega-\varepsilon_F(k))\Gamma^{-1}({\bf k},\omega)-n_0}
\nonumber \\
=&
\frac{(\omega-\varepsilon_F(k))S_1(k)\theta(k_F-k)}{\omega-\Omega_1(k)-i\eta}
\nonumber\\
&+
\frac{(\omega-\varepsilon_F(k))S_1(k)\theta(k-k_F)}{\omega-\Omega_1(k)+i\eta}
\nonumber\\
&+
\frac{(\omega-\varepsilon_F(k))S_2(k)}{\omega-\Omega_2(k)+i\eta}
+
T_c({\bf k},\omega)
\label{tmatrixpoles}
\end{align}
where $T_c({\bf k},\omega)$ is the continuum part and
$(\Omega_i-\varepsilon_F)S_i$ is the residue of the pole at
$\omega = \Omega_i$ (if there is only one pole, we set $S_2=0$).

\begin{figure*}
\includegraphics[width=80mm]{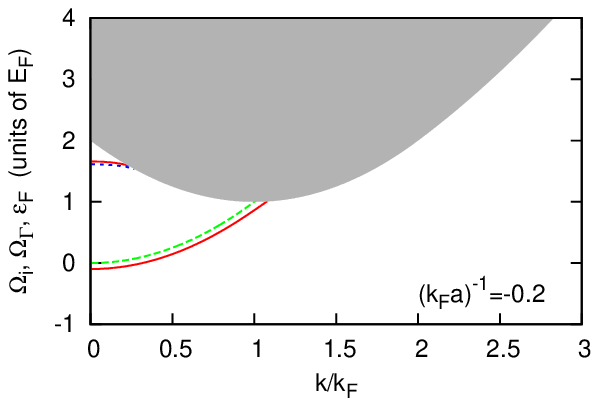}
\includegraphics[width=80mm]{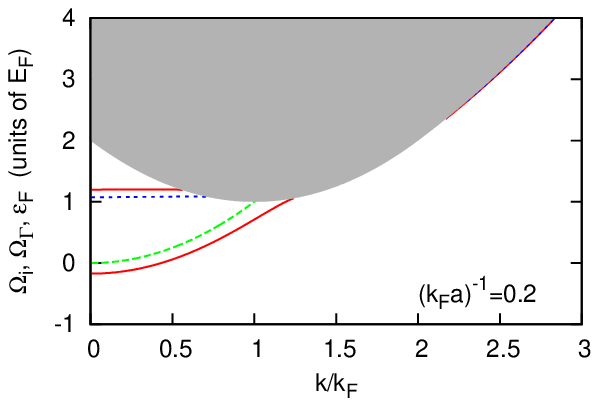}\\
\includegraphics[width=80mm]{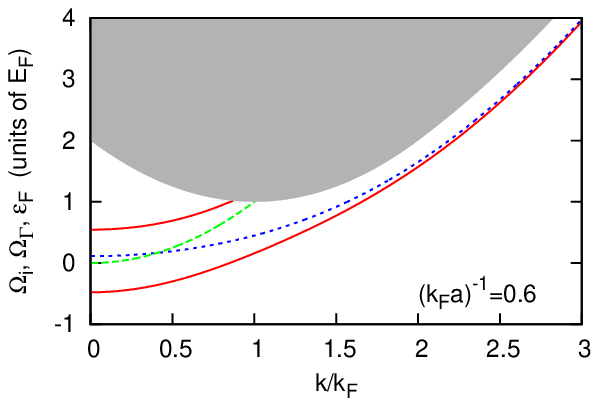}
\includegraphics[width=80mm]{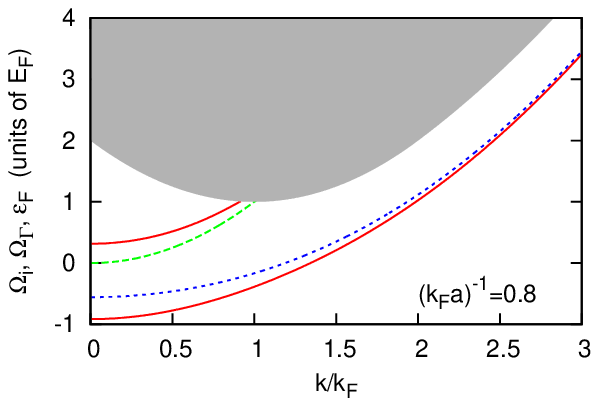}
\caption{\label{fig-dispersion} Dispersion relation of the poles of
  the $T$-matrix for $m_B = m_F$ and $n_B = n_F$ for various values of
  $(k_Fa)^{-1}$. The solid red lines represent poles in
  $T$-matrix. The short-dashed blue line represents the pole in
  $\Gamma$. The long-dashed green line is the non-interacting fermion
  single particle energy $\varepsilon_F(k)=k^2/(2m_F)$. The gray area
  corresponds to the continuum where $\Gamma$ and $T$ have a non-zero
  imaginary part.}
\end{figure*}

In Fig.~\ref{fig-dispersion}
we show the dispersion of the two poles of $T$ for some cases. The
poles are physically of very different nature. $\Omega_2$ is a
collective pole created by BF scattering with the boson always out of
the condensate. $\Omega_1$ stems from the elastic scattering of the
fermion off the Bose condensate and, thus, it is essentially given by
the free fermion dispersion. This is also the reason why $\Omega_1$
lies in the upper half of the complex plane for $k<k_F$, as it is in
the case of the free fermion GF (\ref{Gfermifree}). Of course, in $T$
both branches interact and depending on the system parameters they can
be more or less repelled from one another. An interesting aspect,
already revealed in Ref. \cite{Storozhenko2005}, is that the
$\Omega_2$ branch corresponds to a stable BF pair that exists even for
very weak attraction so that there is no bound state in free
space. This phenomenon is similar to the existence of the Cooper pole
in a pure two species Fermi gas, since the stability of the BF pair in
weak coupling is due to the fact that there is still a sharp Fermi
edge in the problem.

At this point it is worth discussing a subtle point of the theory
related to a possible crossing or inversion of the two branches shown
as the dashed lines in Fig.~\ref{fig-dispersion}. In an unpolarized
spin-1/2 Fermi system, it is known from the Thouless criterion that
once the $T$-matrix has a pole at $\omega = 2\mu$ (where $\mu$ is the
fermion chemical potential), an instability towards the superfluid
(superconducting) state appears, and that for lower temperatures and
in particular for zero temperature the ground state of the system has
to be changed from the Hartree-Fock (HF) to the BCS one \cite{RS}. In
our BF case, one would think that there should be also some criterion
that tells us when our description of a single BF pair in an
uncorrelated ground state becomes invalid and the ground state has to
be changed into a state consisting of many interacting BF pairs. We
are not aware that such a criterion has been given in the
literature. However, we will see in Section~\ref{sec:1p} that as soon
as $\Omega_\Gamma$ drops below $\varepsilon_F$ for $k<k_F$, the
correlation energy does no longer vanish in the limit $n_0\to 0$, as
it should. We therefore suspect that in this case our theory is not
valid any more and we discard in the present work cases in the
parameter space where this happens (e.g., lower panels of
Fig. \ref{fig-dispersion}).

Contrary to the case of spin-1/2 fermions, where the new ground state
of Bose condensed Cooper pairs can be described within BCS theory, it
is unclear how this new ground state of correlated BF pairs should
look like. In any case, as it was pointed out in \cite{Dukelsky2011},
it is obvious that since the BF pairs are fermions, this cannot be
treated as in BCS theory as suggested in \cite{Song2010}. This problem
shall be a very interesting subject for future studies.

\section{Single-particle Green's functions and correlated ground state}
\label{sec:1p}
In order to obtain the occupation numbers in the correlated ground
state, let us get back to the single-particle GFs. As in the standard
NSR approach \cite{NSR}, where the particle number is obtained from a
GF in which the self-energy in the Dyson equation is treated only to
lowest order [diagrams Fig.~\ref{fig-feynmandiagram}(d) and (e)], we
will write for the boson and fermion GFs
\begin{equation}
G_{B,F}({\bf k},\omega)
=
G_{B,F}^0({\bf k},\omega)
+
G_{B,F}^{0\,2}({\bf k},\omega)
\Sigma_{B,F}({\bf k},\omega)
\,.\label{Gfirst}
\end{equation}
The self-energies in ladder approximation are defined by
\begin{multline}
\Sigma_F({\bf k},\omega)
= n_0T({\bf k},\omega)\\
+
i\int \frac{d^3K}{(2\pi)^3}
\int \frac{d\omega'}{2\pi}e^{i\omega'\eta}
T({\bf K},\omega')
G_B^0({\bf K}-{\bf k},\omega'-\omega)\,,
\end{multline}
and
\begin{multline}
\Sigma_B({\bf k},\omega)
=\\
-i\int \frac{d^3K}{(2\pi)^3}
\int \frac{d\omega'}{2\pi}e^{i\omega'\eta}
T({\bf K},\omega')
G_F^0({\bf K}-{\bf k},\omega'-\omega)\,,
\end{multline}
see Feynman diagrams in Fig.~\ref{fig-feynmandiagram}(f) and (g).

By truncating the Dyson equation already at first order in $\Sigma$ in
Eq.~(\ref{Gfirst}), we treat the correlation effects only to leading
order. To be consistent, we should therefore not include the
condensate depletion into the calculation of the $T$ matrix. In other
words, for the condensate density $n_0$ that enters the calculation of
$T$ and $\Sigma_{B,F}$, we put
\begin{equation}
n_0 = n_B\,,
\end{equation}
$n_B$ being the total boson density, since in an uncorrelated system
at zero temperature all bosons are condensed. Although one might be
tempted to use the ``final'' condensate density as a better
approximation for $n_0$, one should remember that standard RPA
\cite{FW,RS} is always built on top of the
uncorrelated ground state, and only in this way one can be sure that
it respects all theorems (see discussion below).

Notice that the first term of $\Sigma_F$ contains one-particle
reducible diagrams (i.e., diagrams that can be separated by cutting a
single fermion line), because in the $T$ matrix the boson can
disappear in the condensate. Nevertheless, this term has to be
retained within RPA, and as long as $\Sigma_F$ is kept only to first
order in Eq.~(\ref{Gfirst}) this is not a problem.

Using the above equations, the fermion and boson occupation numbers
can be calculated from
\begin{equation}
\rho_{B,F}(k)
=
\pm i\int\frac{d\omega}{2\pi}e^{i\eta\omega}G_{B,F}({\bf k},\omega)
\label{occgeneral}\,,
\end{equation}
the upper (lower) sign being valid for bosons (fermions). Inserting
the explicit expressions for the self-energies, one obtains
\begin{widetext}
\begin{gather}
\rho_F(k)
=
\theta(k_F-k)\Gamma^{-1}(k,\Omega_1(k))S_1(k)
+\theta(k-k_F)
\int\frac{d^3K}{(2\pi)^3}
\frac{(\Omega_1(K)-\varepsilon_F(K))S_1(K)\theta(k_F-K)}
{[\Omega_1(K)-\varepsilon_F(k)-\varepsilon_{B}({\bf K}-{\bf k})]^2}\,,
  \label{fermiocc}\\
\rho_B(k)
=
\int \frac{d^3K}{(2\pi)^3}S_1(K)(\Omega_1(K)-\varepsilon_F(K))
\frac{\theta(k_F-K)\theta(|{\bf K}-{\bf k}|-k_F)}
{[\Omega_1(K)-\varepsilon_F({\bf K}-{\bf k})-\varepsilon_B(k)]^2}\,.
\end{gather}
\end{widetext}

\begin{figure}
\includegraphics[width=80mm]{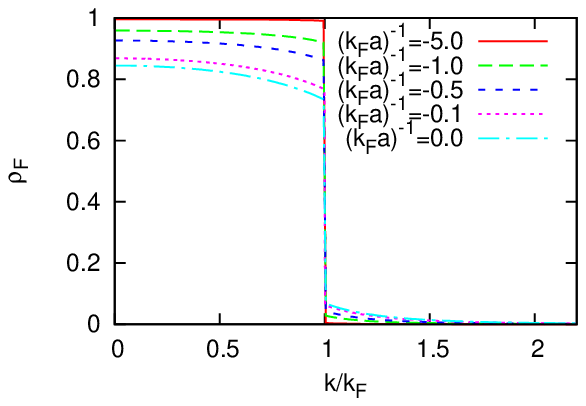}
\includegraphics[width=80mm]{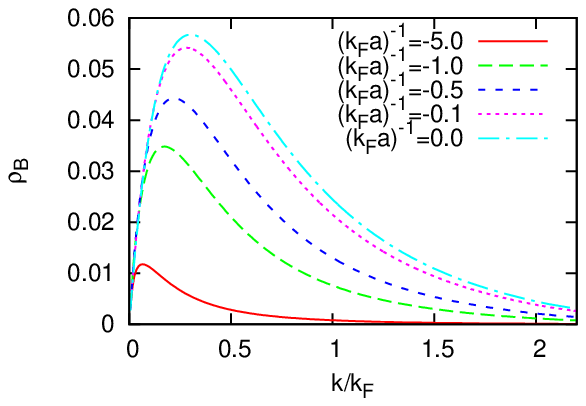}
\caption{\label{fig-occupationnumber} Occupation numbers of fermions
  and bosons at $m_B = m_F$ and $n_B = n_F$ for various values of
  $(k_Fa)^{-1}$.}
\end{figure}

The results for the occupation numbers are presented in
Fig.~\ref{fig-occupationnumber}
for various system parameters. Note that, as a consequence of the
perturbative treatment of the self-energy in Eq. (\ref{Gfirst}), the
$Z$ factor determining the jump of $\rho_F$ at the Fermi surface can
become negative, or the number of bosons out of the condensate can
become larger than the total number of bosons. We discard such cases
and restrict ourselves to parameters in which the correlations are not
too strong.

The Luttinger theorem states that in the fermion distribution the
momentum integral over what is missing with respect to the free case
below $k_F$ is exactly compensated by the part above $k_F$, i.e.,
\begin{equation}
\int_{k<k_F} \frac{d^3k}{(2\pi)^3} (1-\rho_F(k)) = \int_{k>k_F}
\frac{d^3k}{(2\pi)^3} \rho_F(k)\,.
\label{luttingertheorem}
\end{equation}
From general properties of RPA theory (see below) one expects that the
Luttinger theorem should be exactly fulfilled in our scheme, although
from the final expression (\ref{fermiocc}) for the occupation numbers
this is hard to see. In our numerical calculations,
Eq. (\ref{luttingertheorem}) is fulfilled to a relative accuracy of
better than $10^{-3}$. This is the advantage of treating the
self-energy perturbatively in Eq. (\ref{Gfirst}). If we had resummed
the Dyson equation to all orders, as in
\cite{Fratini2010,Fratini2012,Fratini2013}, the Luttinger theorem
would most likely have been violated. For instance, in Fig. 8(a) of
Ref.~\cite{Fratini2012} it seems that the number of fermions above the
Fermi surface\footnote{Note that in Ref.~\cite{Fratini2012} the Fermi
  surface is not at $k=k_F$ because $k_F$ has a different meaning in
  that paper.} is larger than the number of fermions missing below.

\begin{figure}
\includegraphics[width=80mm]{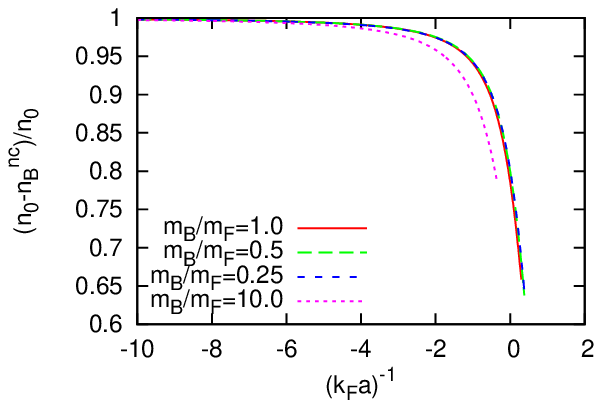}
\includegraphics[width=80mm]{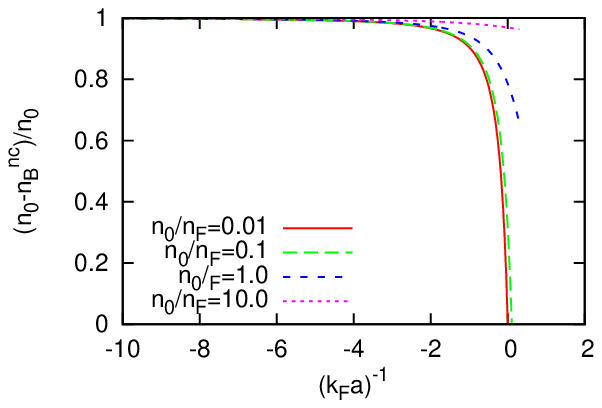}
\caption{\label{fig-depletion} Condensate depletion as a function of
  the interaction strength: (a) for various mass ratios $m_B/m_F$ and
  $n_B = n_F$; (b) for various density ratios $n_B/n_F$ and $m_B =
  m_F$.}
\end{figure}

In addition to the Luttinger theorem (\ref{luttingertheorem}) for the
fermions, our formulation satisfies the following relation:
\begin{gather}
n_B^{nc} = \int_{k>0} \frac{d^3k}{(2\pi)^3} \rho_B(k) = 
  \int_{k>k_F}\frac{d^3k}{(2\pi)^3} \rho_F(k)\,.
\label{depletion}
\end{gather}
where $n_B^{nc}$ denotes the density of non-condensed bosons. The
relation has a very intuitive interpretation: each time a boson is
scattered out of the condensate, also a fermion is scattered out of
the Fermi sea. Therefore the total number of fermions above $k_F$ must
be equal to the number of bosons out of the condensate. The condensate
depletion as a function of the interaction strength is shown in Fig.\@
\ref{fig-depletion} for different mass and density ratios. As
mentioned before, we stop the calculation as soon as $n_0-n_B^{nc}$ or
the $Z$ factor of the fermions becomes negative or the branch
$\Omega_\Gamma$ drops below $\varepsilon_F$ for $k<k_F$.

Let us discuss the approximation scheme which is set up in the
foregoing equations. One recognizes the similarity with the NSR
approach for the treatment of interacting two-component Fermi gases
close to the transition temperature to the superfluid state
\cite{NSR}. The main difference is that the NSR formalism is
transcribed here to an interacting BF system at zero temperature.

The $T$-matrix, which sums in the case of a pure Fermi system
simultaneously the particle-particle (pp) and hole-hole (hh) ladders,
is sometimes also called the pp-RPA \cite{RS}. It is well known that
RPA theory has appreciable properties as the fullfillment of
conservation laws and sumrules. (The latter statements are,
strictu-senso, only valid if the RPA is solved in the HF basis
\cite{RS}. However, in our case the HF shifts are unimportant because
they disappear in the regularization procedure when the coupling
constant $g$ tends to zero while the cut-off tends to infinity,
keeping the scattering length $a$ constant
\cite{Watanabe2008,Fratini2010,Fratini2012}.)

It is, in principle, straight forward to prove that the Luttinger
theorem is fulfilled in strict application of RPA. The proof is
straight-forward and well known in the case of particle-hole (ph) RPA
in a system with discrete single particle states as it is often
considered in, e.g., nuclear or atomic and chemical physics
\cite{RS,Rowe,Casida}, i.e., for finite Fermi systems\footnote{In
  \cite{Rowe,Bouyssy1974} explicit expressions for the correlated
  parts of the single particle occupation numbers are given. From
  these expressions, it becomes so obvious that particle number is
  conserved that this property is most of the time not even stated in
  the literature.}. In the case of pp-RPA, things are less well known
but corresponding expressions can also be found in the literature
\cite{Bouyssy1974}. In the BF case, the fulfillment of the Luttinger
theorem, i.e., the fact that the occupation numbers of levels above
the Fermi surface exactly cancel the reduction of the occupation
numbers of levels below the Fermi surface, has also been demonstrated
for finite size cases with the Hubbard model \cite{Barillier2008}. To
our knowledge, it has never been shown with RPA in continuum cases
where things are, of course, a little more tricky, for instance from
the numerical point of view.

It is, however, very important to notice a subtle difference between
this strict RPA approach and the NSR scheme. The latter is generally
formulated in finite-temperature formalism, and the zero-temperature
case is obtained as a limiting procedure as, e.g., in
\cite{Fratini2012}. However, the two formalisms do not become
equivalent in this limit (see, e.g., chapter 3.3 of
\cite{NegeleOrland}), even if the self energy is only treated to first
order and not summed as in \cite{Fratini2012}. We will elaborate in a
forthcoming paper on this point.

In our scheme, the particle numbers $n_{B,F}$ are fixed from the
beginning and they are not modified by the inclusion of correlations
(because the Luttinger theorem is satisfied). Therefore we cannot
determine the chemical potentials in the way this is usual done in the
NSR scheme by inverting the $n(\mu)$ relation obtained by integrating
Eq.\@ (\ref{occgeneral}) over $k$. But of course, also in our scheme
the correlations change the equation of state, i.e., the relation
between $n$ and $\mu$. Therefore, we have free chemical potentials,
$\mu_F^0 = E_F = k_F^2/(2m_F)$ and $\mu_B^0 = 0$, and modified ones
$\mu_{F,B}$ that include the correlation effects. But here the
corrections to the chemical potentials are calculated perturbatively
to first order in the correlations. They are obtained from the
correlated ground state energy density, i.e.,
\begin{equation}
\mu_{F,B}= \frac{\partial \mathcal E}{\partial n_{F,B}}\,.
\end{equation}
\noindent
The energy density $\mathcal E$ is calculated within RPA in the usual
way from the coupling constant integration \cite{FW}
\begin{equation}
\label{correlationenergy}
\mathcal E -\mathcal E_0
= 
-i\int_0^1\frac{d\lambda}{\lambda}
\int \frac{d^3k}{(2\pi)^3}
\int \frac{d\omega}{2\pi}
e^{i\omega \eta}\Sigma_F^{\lambda}(k,\omega)G^0_F(k,\omega)\,,
\end{equation}
where $\Sigma_F^\lambda$ is the self-energy calculated with coupling
constant $g\lambda$ instead of $g$. Considering a finite value of the
coupling constant $g$ and a cutoff and taking the cutoff to infinity
only in the end of the calculation, one obtains the following simple
formula for the ground-state energy:
\begin{equation}
\mathcal E -\mathcal E_0 =
\int_{k<k_F}\frac{d^3k}{(2\pi)^3}[\Omega_1(k) -\varepsilon_F(k)]\,.
\end{equation}
This expression for the energy density agrees with that given in
\cite{Storozhenko2005} besides the fact that here the extra term of
the fermion-hole boson-condensate matrix element is missing, since it
has been absorbed by the regularization procedure. From this formula
it is clear that as long as $\Omega_\Gamma$ lies above $\varepsilon_F$
for $k<k_F$, the branch $\Omega_1$ approaches $\varepsilon_F$ in the
limit $n_0\to 0$ and the correlation energy tends to zero, which is
not true if $\Omega_\Gamma$ drops below $\varepsilon_F$ for $k<k_F$.

\begin{figure}
\includegraphics[width=80mm]{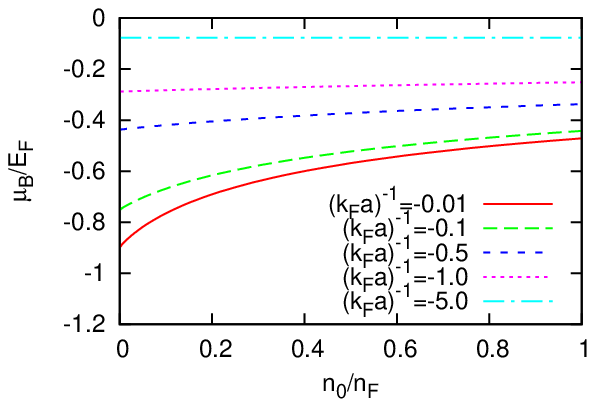}
\includegraphics[width=80mm]{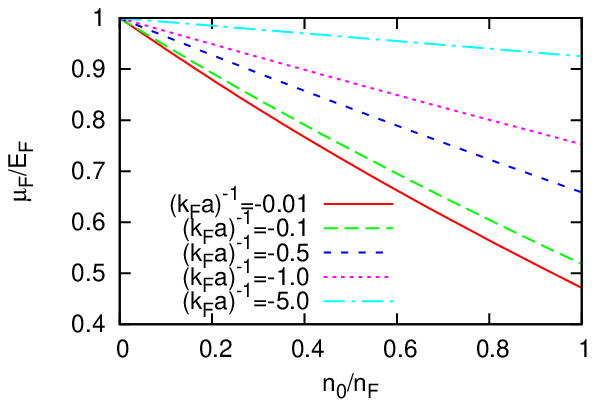}
\caption{\label{fig-chemicalpotential} Fermion and boson chemical
  potentials as a function of the boson density $n_0$ for various
  values of the scattering length and $m_B=m_F$.}
\end{figure}

The boson and fermion chemical potentials calculated in this way are
shown in Fig.~\ref{fig-chemicalpotential}. As expected, we see that
the chemical potentials are lowered by the correlations if the boson
density $n_0$ or the interaction strength $|a|$ ($a<0$) increases.
One can show analytically that the boson chemical potential satisfies
\begin{equation}
\mu_B = \Sigma_B(0,0)\,,
\end{equation}
which is the usual condition for Bose condensed systems.

\begin{figure}
\includegraphics[width=80mm]{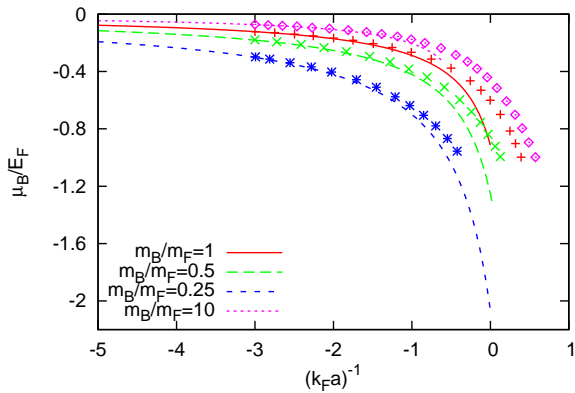}
\caption{\label{fig-polaronlimit}Boson chemical potential as a
  function of $(k_Fa)^{-1}$ at $n_0/n_F=0.001$ for various mass
  ratios. The symbols $+$, $\times$, $\ast$, and $\diamond$ are
  polaron chemical potentials extracted from Fig.\@ 1 of Ref.\@
  \cite{Combescot2007} for $m_B/m_F = 1$, $0.5$, $0.25$, and
  $\infty$.}
\end{figure}

Let us now consider the case with almost vanishing boson number, i.e.,
the polaron limit, where it is immaterial whether the impurity is a
boson or a fermion of another species (or opposite spin). Boson and
fermion chemical potentials in this limit are displayed in
Fig.~\ref{fig-polaronlimit}
as functions of the interaction strength for various mass ratios
$m_B/m_F$. We compare our results with those of Combescot et
al. \cite{Combescot2007}. We see that the agreement is quite good for
negative and not too large scattering lengths. For values of
$(k_Fa)^{-1}$ close to $-1$ the agreement deteriorates. This is not
surprising, since we treat the self-energy only to first order whereas
in the polaron approach the whole series is summed.

It would therefore be desirable to sum up the selfenergy to all
orders. However, with the present form of the selfenergy, this would
cause other problems, such as the violation of the Luttinger theorem
(\ref{luttingertheorem}). We think that these issues should be settled
before definite conclusions can be drawn from a non-perturbative
approach.

\section{Summary, Discussion, and Outlook}
\label{sec:summary}
In this work we used a $T$ matrix approach to describe BF pair
correlations in a BF mixture. The approach is very similar to the
usual NSR theory for fermions \cite{NSR}. However, there are subtle
differences because we work within the zero-temperature formalism. Our
approach is a strict application of what has been known as pp-RPA in
nuclear physics \cite{RS,Bouyssy1974}.  As expected, this approach
respects the Luttinger theorem. This is explicitly verified
numerically to high precision in calculating the correlated fermion
and boson occupation numbers. We also studied for the bosons the
condensate depletion and found that the number of bosons scattered out
of the condensate is exactly equal to the number of fermions scattered
above the Fermi surface. In studies of spin-1/2 Fermi gases, it is
often supposed that the Luttinger theorem is satisfied (see, e.g.,
Eq. (6) of Ref. \cite{Navon2013}) but it is rarely checked whether the
approximations that are used preserve this property. The problems
found in studies of polarized Fermi systems
\cite{LiuHu2006,Parish2007,Kashimura2012a,Kashimura2012b} might also
be related to this problem.

As in the original NSR approach, we keep the self-energies only to
first order in the Dyson equation. Besides the nice properties
mentioned before, this has of course also some drawbacks. For
instance, the $Z$ factor of the fermion GF (i.e., the jump of the
occupation numbers at $k_F$) may become negative if the correlations
are too strong. A possible way to avoid this over-estimation of the
correlation effects, without violating the Luttinger theorem, would be
to use in the $T$ matrix the self-consistently determined correlated
occupation numbers instead of the uncorrelated ones. In nuclear
physics this approximation is known as ``renormalized RPA'', see, e.g.
\cite{Hirsch2002,Delion2005}.

We also investigated the polaron limit and found that the boson
chemical potential agrees well with the results by Combescot et
al. \cite{Combescot2007} in the weak-coupling region. Close to
unitarity the results start to diverge, which is again a consequence
of our perturbative treatment of the self-energy.

If one goes in the molecular regime beyond the polaron limit, one
expects the system to have a completely different ground state, namely
a Fermi sea of composite molecules. Actually this transition might
already happen before the molecular limit, since there is, as in the
Cooper pair problem, always a stable BF branch in the in-medium $T$
matrix, even if in free space there is no bound state. How this
transition happens is still unclear \cite{Dukelsky2011} and needs
further investigation.

\acknowledgments
Discussions with R. Combescot, X. Leyronas, and P. Pieri are
gratefully acknowledged.

\end{document}